# Gastro-Intestinal Tract Segmentation Using an Explainable 3D Unet

Kai Li
kaii.li@mail.utoronto.ca
University of Toronto
Toronto, Ontario, Canada

Jonathan H. Chan
jonathan@sit.kmutt.ac.th
King Mongkut's University of Technology Thonburi
Bangkok, Thailand

**ABSTRACT**

In treating gastrointestinal cancer using radiotherapy, the role of the radiation oncologist is to administer high doses of radiation, through x-ray beams, toward the tumor while avoiding the stomach and intestines. With the advent of precise radiation treatment technology such as the MR-Linac, oncologists can visualize the daily positions of the tumors and intestines, which may vary day to day. Before delivering radiation, radio oncologists must manually outline the position of the gastrointestinal organs in order to determine position and direction of the x-ray beam. This is a time consuming and labor intensive process that may substantially prolong a patient's treatment. A deep learning (DL) method can automate and expedite the process. However, many deep neural networks approaches currently in use are black-boxes which lack interpretability which render them untrustworthy and impractical in a healthcare setting. To address this, an emergent field of AI known as Explainable AI (XAI) may be incorporated to improve the transparency and viability of a model. This paper proposes a deep learning pipeline that incorporates XAI to address the challenges of organ segmentation.

## 1 INTRODUCTION

In 2019, an estimated 5 million people worldwide were diagnosed with cancer of the gastrointestinal tract. About half of these patients are eligible for radiation therapy (RT), a form of treatment that can provide palliative relief, if not improve the rates of cure [1]. RT is delivered for 10 to 15 minutes a day for 1 to 6 weeks, which requires radiation oncologists to frequently label the position of the stomach and intestines in order to avoid them during treatment. This paper discusses a DL pipeline that can generate inferences for MRI volumes, while also leveraging XAI to create heatmaps for interpretation [2]. The pipeline consists of a 3D UNet Model that applies the following XAI techniques as post processing: GradCAM[3][4], guided GradCAM[5], and DeepLifT[6]. These post processing techniques are useful in demonstrating how various spatial regions of the input influence the prediction of a particular class; they identify the salient features that lead to the model's prediction. In terms of accuracy of the model, it is able to achieve a testing accuracy of 86.5% through ensembling multiple training folds.

## 2 RELATED WORKS

Towards Interpretable Semantic Segmentation via Gradient-weighted Class Activation Mapping [15] is a paper by Vinogradova et al. that proposes an XAI SEG-GRAD-CAM, a gradient based method for image segmentation that extends from the widely-used GRAD-CAM commonly employed in classification tasks. Whereas classification networks output a single class distribution per image, $y^c$, segmentation networks produce logits for each image pixel of a class, which is denoted as $y^c_{i,j}$. Accordingly, in SEG-GRAD-CAM, $y^c$ in GRADCAM is replaced with $\sum_{x,y \in M} y^c_{i,j}$ where $M$ denotes a set of pixel indices of interest. During experiments, SEG-GRAD-CAM was applied to various convolutional layers in the encoder of a UNet that was trained on the Cityscapes dataset. As expected, heatmaps in initial layers exhibit edge-like structures which is consistent with the fact that early layers discern low level features. Applying SEG-GRAD-CAM at the end of the contracting path—the bottleneck layer—highlighted regions of the input that show the abstract features contributing to a certain class. For instance, the pixels belonging to trees are highlighted for the prediction of the class 'sky', indicating these are informative for the prediction of 'sky' pixels.

## 3 METHODOLOGY

### 3.1 Datasets

The dataset used to train the model consisted of 300 image volumes and their corresponding masks rendered from 2D MRI scans. These MRI scans are from actual cancer patients who had 1-5 scans on separate days during their treatment. These scans vary significantly in size. There are 4 unique sizes: 234 x 234, 266 x 266, 276 x 276, and 310 x 360 pixels. Each scan has a corresponding mask segmenting the small bowel, large bowel, and stomach. There are several instances where masks overlap one another, which makes this a multilabel segmentation task. To improve the prediction accuracy, a 5 fold ensemble training approach [7] was used. Each fold uses approximately 240 volumes for training—80% of the dataset, and the remaining 60 for validation–20% of the dataset. The stratified group K fold method is used to ensure folds are balanced. Namely, each fold contains the same proportion of annotations for each volume. For testing, the model is run against a hidden test set that consists of about 50 cases, where each case contains anywhere from 1 to 5 volumes. The training and testing images are available on Kaggle's "UW-Madison GI Tract Image Segmentation" competition [1].

## 3.2 Image Preprocessing

Each input is first cropped in order to remove unnecessary background space, then normalized. The model is trained on smaller patches of 160 x 160 x 80 pixels rather than an entire volume, primarily because of memory limitations of computing resources. These patches are created using a 3D random spatial crop. Other transformations include random flip, random affine, random grid distortion, random dropouts, and random shifting and scaling in intensity—each with a probability value of 0.5. These help reduce the chances of overfitting.

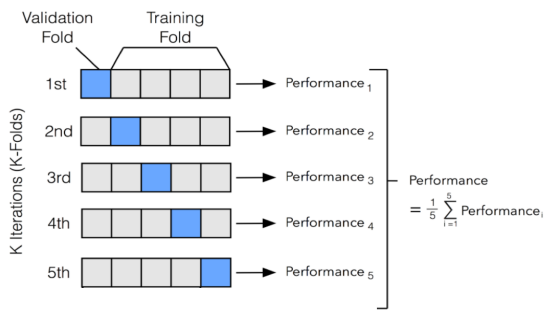

**Figure 1: Illustration of 5 fold ensembling. Ensembled model has better generalization performance compared to each individual fold**

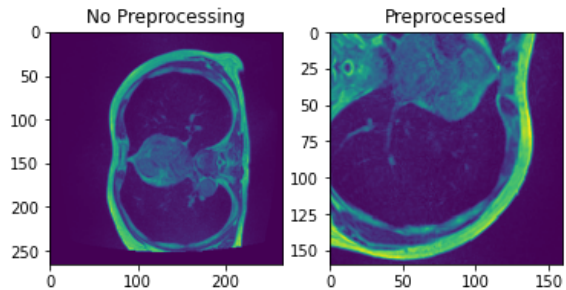

**Figure 2: Sample raw image compared to preprocessed image, which features a random crop, change in intensity, and random flip.**

## 3.3 Model

The model is implemented on Pytorch, and uses MONAI's 3D UNet architecture. The architecture consists of 3 sections: A contraction which produces feature maps, a bottleneck which mediates between the downsampling and upsampling paths, and an expansion section which restores the maps back into the original size [8]. There are 5 layers of contraction and expansion path. In the contraction, each layer applies a 3D convolution, normalization, dropout, and activation function. The normalization used is BatchNorm, the dropout value is 0.2, and the activation function is PReLU—Parametric Rectified Linear Unit.

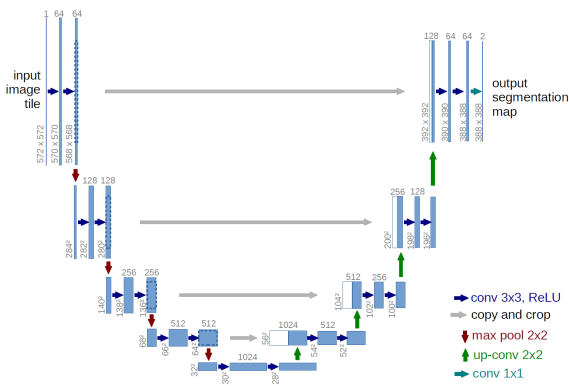

**Figure 3: UNet Model as described by original paper. Note the activation function for the model in this paper is PReLU rather than ReLU**

The AdamW optimizer [9] was used with an initial learning rate of $5 \times 10^{-4}$, which decays to $1 \times 10^{-4}$ through a learning rate scheduler, which is a one cycle learning rate [10]. Each fold is trained on 120 epochs initially, then fine tuned using 40 epochs on a lower learning rate of $3 \times 10^{-4}$. Due to memory limitations on the machine, a batch size of 4 was chosen, although it is not necessarily optimal. For the same reason, the number of epochs is also limited.

The predictions of this model were submitted to Kaggle's "UW-Madison GI Tract Image Segmentation" competition for testing.

The final accuracy of the model was to be tested using a specific metric imposed by the competition:

$0.4 \times DSC + 0.6 \times hausdorff\ distance$

DSC is the dice similarity coefficient [11], given as:

$DSC = \frac{2 \times |X \cap Y|}{|X| + |Y|}$

The Hausdorff distance [11] is calculated as follows:

$d_H(X, Y) = \max \left\{ \sup_{x \in X} d(x, Y), \sup_{y \in Y} d(X, y) \right\}$

In the above equations, X and Y denote the tensors of the predicted mask and the ground truth mask.

| Parameter | Run 1 | Run 2 |
|---|---|---|
| Batch Size | 4 | 4 |
| Initial Learning Rate | $5 \times 10^{-4}$ | $3 \times 10^{-4}$ |
| Epochs | 120 | 40 |
| Minimum Learning Rate | $1 \times 10^{-4}$ | $1 \times 10^{-4}$ |

**Table 1: Table summarizing various hyperparameters for the two training runs used to train each fold. The initial learning rate was set to ensure stable training yet quick convergence due to the limited number of epochs.**

Although the testing metric is a linear combination of both DSC and Hausdorff distance, in computer vision literature, segmentation models generally do not attempt to directly minimize Hausdorff distance due to its sensitivity to noise and outliers. Moreover, Hausdorff distance is determined solely by the largest error, and using it as a loss function may lead to poor segmentation performance and algorithm instability [12]. Therefore, the loss function used for this model was purely dice loss, which is equal to 1 - dice similarity coefficient (DSC), or $1 - \frac{2 \times |X \cap Y|}{|X| + |Y|}$. It is worth noting that cross entropy [b] is another common loss function in medical image segmentation.

Although DSC generally leads to better results [12], it is also possible to utilize a weighted loss function combining the two metrics. This may in fact help the model learn the right features better which could result in a higher test score.

### 3.4 Visualizations

For model interpretability, two pytorch visualization libraries were used. Namely, *Pytorch GradCAM* [13] was used for gradient based methods, which includes the GradCAM and guided GradCAM implementation, while *Captum* [14] was used for the activation based method, DeepLifT.

### 4 RESULTS

The visualizations generated from GradCAM, DeepLifT and Guided GradCAM are shown in figures 4, 5, and 6 respectively. Each visualization is shown next to its ground truth mask. Training results for each fold are displayed in table 2. Note that total accuracy is calculated using the weighted average of the validation DSC and validation hausdorff distance, as specified in section **3.3**.

| Fold | Training DSC | Validation DSC | Validation Hausdorrf distance | Total Accuracy |
|---|---|---|---|---|
| Fold 1 | 0.865 | 0.825 | 0.962 | 0.906 |
| Fold 2 | 0.868 | 0.805 | 0.960 | 0.898 |
| Fold 3 | 0.860 | 0.812 | 0.963 | 0.903 |
| Fold 4 | 0.903 | 0.820 | 0.964 | 0.906 |
| Fold 5 | 0.833 | 0.825 | 0.943 | 0.899 |

**Table 2: Final training DSC, validation DSC, validation Hausdorff distance, and total score for each training fold.**

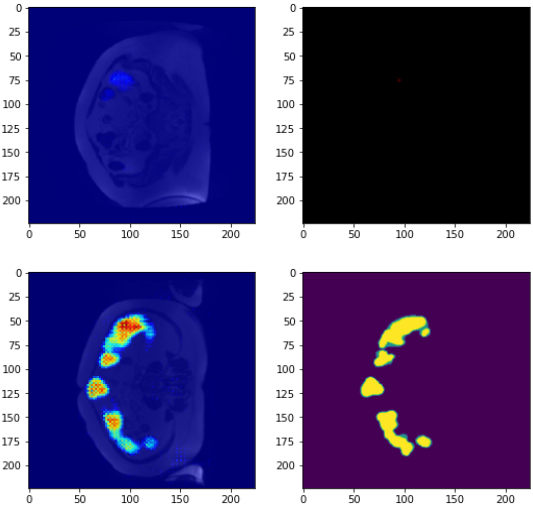

**Figure 4 : GradCAM visualization showing areas contributing to the prediction of class "small bowel" compared to its ground truth mask. Warmer regions indicate greater activation; These have more influence on the model's decision.**

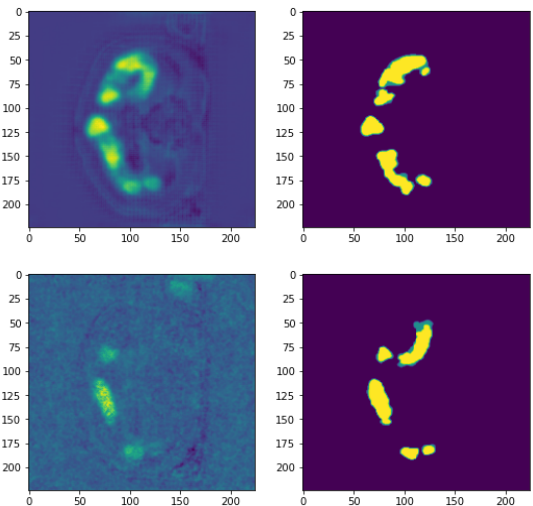

**Figure 5: Deeplift visualization for the activation of class "small bowel" compared to ground truth mask. Light green highlights indicate regions of high relevance, faded green indicates less discriminative regions**

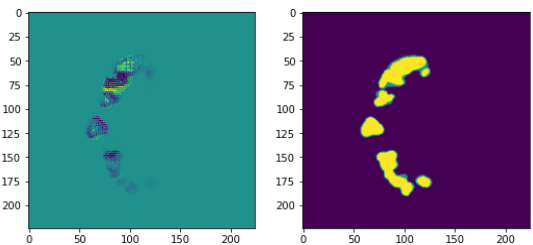

**Figure 6: Guided GradCAM visualization for the activation of class "small bowel" compared to its ground truth mask. Guided GradCAM incorporates guided backpropagation which also identifies negative gradients.**

## 5 DISCUSSION

The ensembled 3D UNet is a strong performing model as it is able to achieve a score of 0.855 which is among the top 40% of scores in the competition. Based on the visualizations, there remain several areas that could be improved. For instance, in the DeepLifT heatmaps, the model fails to detect some portions of the organ entirely. Furthermore, there is activation in the upper part of the scan, which is an area of the arm. These errors can be minimized through better preprocessing, such as smarter cropping of the input images. Another observation is that for certain folds there is a substantial discrepancy between the validation score and the training score: in fold 4, the training DSC is 0.903 while the validation DSC is only 0.820. It was also noted that around the last 20 epochs of training, the validation DSC began decreasing with each subsequent epoch despite training DSC increasing. These are indications of overfitting. One solution for this is to increase the data augmentation or transformations—for instance, adding random scaling of contrast and random noise. Visualizations also show there are several improvements that can be made to the training data. One error is that the labels are often truncated so the bottom parts of the volume are unlabelled. This can be seen in figure 4 where visualizations clearly indicate the presence of the small bowel despite the mask being empty.

## 6 CONCLUSION

A vanilla 3D UNet is able to achieve relatively accurate predictions, based on its performance in the competition. This model is intended to be a baseline and its performance can be further enhanced by using transfer learning, that is, applying an advanced classification algorithm such as EfficientNet, ResNet, VGG16, and more, as the encoder for the UNet. Improving faulty training data could also help strengthen the model. As next steps, one could label the bottom slices of the training volume which are missing masks or leave them out of training altogether. One could also experiment with different loss functions, such as Focal Loss, Binary Cross Entropy Loss (BCE Loss), and Intersection over Union (IoU).

## REFERENCES

[1] 'UW-Madison GI Tract Image Segmentation'. Kaggle, 2022.

[2] G. Yang, Q. Ye, and J. Xia, "Unbox the black-box for the medical explainable AI via multi-modal and Multi-centre Data Fusion: A mini-review, two showcases and beyond," *arXiv.org*, 03-Feb-2021. [Online]. Available: https://arxiv.org/abs/2102.01998. [Accessed: 22-Dec-2022].

[3] H. Panwar, P. K. Gupta, M. K. Siddiqui, R. M. Menendez, P. Bhardwaj, and V. Singh, "A deep learning and grad-cam based color visualization approach for fast detection of covid-19 cases using chest X-ray and CT-scan images," *Chaos, Solitons & Fractals*, 07-Aug-2020. [Online]. Available: https://www.sciencedirect.com/science/article/pii/S0960077920305865. [Accessed: 23-Dec-2022].

[4] R. R. Selvaraju, M. Cogswell, A. Das, R. Vedantam, D. Parikh, and D. Batra, "Grad-CAM: Visual explanations from deep networks via gradient-based localization," in Proceedings of the IEEE International Conference on Computer Vision, Oct. 2017, pp. 618-626.

[5] R. R. Selvaraju, A. Abulnaga, M. Cogswell, and D. Batra, "Guided Grad-CAM: Why did you say that? Visual explanations from deep networks through guided backpropagation," in Proceedings of the IEEE Conference on Computer Vision and Pattern Recognition, Jun. 2020, pp. 9364-9372.

[6] R. R. Selvaraju, M. Cogswell, A. Das, R. Vedantam, D. Parikh, and D. Batra, "DeepLiFT: Understanding and improving deep networks by leveraging feature transforms," in Proceedings of the Conference on Neural Information Processing Systems, Dec. 2017, pp. 6078-6087.

[7] Yahia, N. B., Kandara, M. D., & Saoud, N. B. (2020). Deep Ensemble Learning Method to forecast COVID-19 outbreak. https://doi.org/10.21203/rs.3.rs-27216/v1

[8] O. Ronneberger, P. Fischer, and T. Brox, "U-Net: Convolutional networks for biomedical image segmentation," in Proceedings of the International Conference on Medical Image Computing and Computer-Assisted Intervention, Sep. 2015, pp. 234-241.

[9] I. Loshchilov and F. Hutter, "Decoupled weight decay regularization," in Proceedings of the International Conference on Learning Representations, May 2019.

[9] H. Gelband, P. Jha, R. Sankaranarayanan, S. Horton, W. Bank, D. A. Jaffray, and M. K. Gospodarowicz, "Radiation Therapy for Cancer," in *Disease control priorities, third edition (volume 3): Cancer*, Washington: World Bank Publications, 2015, pp. 239–240.

[10] L. N. Smith, "Cyclical learning rates for training neural networks," *2017 IEEE Winter Conference on Applications of Computer Vision (WACV)*, 2015. doi:10.1109/wacv.2017.58

[11] A. A. Taha and A. Hanbury, "Metrics for evaluating 3D medical image segmentation: Analysis, selection, and tool," *BMC Medical Imaging*, vol. 15, no. 1, 2015.

[12] D. Karimi and S. E. Salcudean, "Reducing the Hausdorff distance in medical image segmentation with Convolutional Neural Networks," *IEEE Transactions on Medical Imaging*, vol. 39, no. 2, pp. 499–513, Apr. 2019. doi:10.1109/tmi.2019.2930068

[13] J. Gildenblat and contributors, "PyTorch library for CAM methods," GitHub, 2021, [Online]. Available: https://github.com/jacobgil/pytorch-grad-cam.

[14] N. Kokhlikyan, V. Miglani, M. Martin, E. Wang, B. Alsallakh, J. Reynolds, A. Melnikov, N. Kliushkina, C. Araya, S. Yan, and O. Reblitz-Richardson, "Captum: A unified and generic model interpretability library for PyTorch," in arXiv preprint arXiv:2009.07896, cs.LG, Sep. 2020.

[15] K. Vinogradova, A. Dibrov, and G. Myers, "Towards interpretable semantic segmentation via gradient-weighted class activation mapping (student abstract)," *Proceedings of the AAAI Conference on Artificial Intelligence*, vol. 34, no. 10, pp. 13943–13944, 2020. doi:10.1609/aaai.v34i10.7244